\documentclass[12pt]{article}
\pdfoutput=1

 \usepackage{amsmath,amsfonts,amssymb,mathrsfs,graphicx,xcolor,ifthen,times}
\usepackage{scicite}
\usepackage{hyperref}
\usepackage[normalem]{ulem}

\newenvironment{sciabstract}{%
\begin{quote} \bf}
{\end{quote}}
\newcounter{lastnote}

\newcommand{\fig}{Fig.}
\newcommand{\efig}{Supplementary Figure}
\newcommand{\eTab}{Supplementary Table}

\newcommand{\figu}[1]{\fig~\ref{fig:#1}}

\newcommand{\efigu}[1]{\efig~\ref{fig:#1}}

\newcommand{\bi}{\begin{itemize}}
\newcommand{\ei}{\end{itemize}}

\newcounter{nref}
\newcommand{\apj}{Astrophys. J.}
\newcommand{\pasp}{PASP} 

\title{Modelling the coincident observation of a high-energy neutrino and a bright blazar flare}  

\author
{Shan Gao$^1$, Anatoli Fedynitch$^1$, Walter Winter$^1$, and  Martin Pohl$^{1,2}$ \\
\\
\normalsize{$^1$Deutsches Elektronen-Synchrotron (DESY), Platanenallee 6, D-15738 Zeuthen, Germany} \\
\normalsize{$^2$Institute of Physics and Astronomy, University of Potsdam, D-14476 Potsdam, Germany}\\
}

\date{\today}

\begin{document}

\baselineskip24pt

\maketitle 

\begin{sciabstract}

In September 2017, the IceCube Neutrino Observatory recorded a very-high-energy neutrino in directional coincidence with a blazar in an unusually bright gamma-ray state,  TXS0506+056  \cite{IceCube_GCN,TXS_MM}. Blazars are prominent photon sources in the universe because they harbor a relativistic jet whose radiation is strongly collimated and amplified. High-energy atomic nuclei known as cosmic rays can produce neutrinos; thus the recent detection may help identifying the sources of the diffuse neutrino flux \cite{Aartsen:2016xlq} and the energetic cosmic rays. Here we report on a self-consistent analysis of the physical relation between the observed neutrino and the blazar, in particular the time evolution and spectral behavior of neutrino and photon emission. We demonstrate that a moderate enhancement in the number of cosmic rays during the flare can yield a very strong increase of the neutrino flux which is limited by co-produced hard X-rays and TeV gamma rays. We also test typical radiation models \cite{1995PASP..107..803U,2013ApJ...768...54B} for compatibility and identify several model classes \cite{1992AA...253L..21M, 2001APh....15..121M} as incompatible with the observations. We investigate to what degree the findings can be generalized to the entire population of blazars, to determine the relation between their output in photons, neutrinos, and cosmic rays, and suggest how to optimize the strategy of future observations.
\end{sciabstract}

\maketitle

\begin{figure}[t]
\begin{center}
\includegraphics[width=0.75\textwidth]{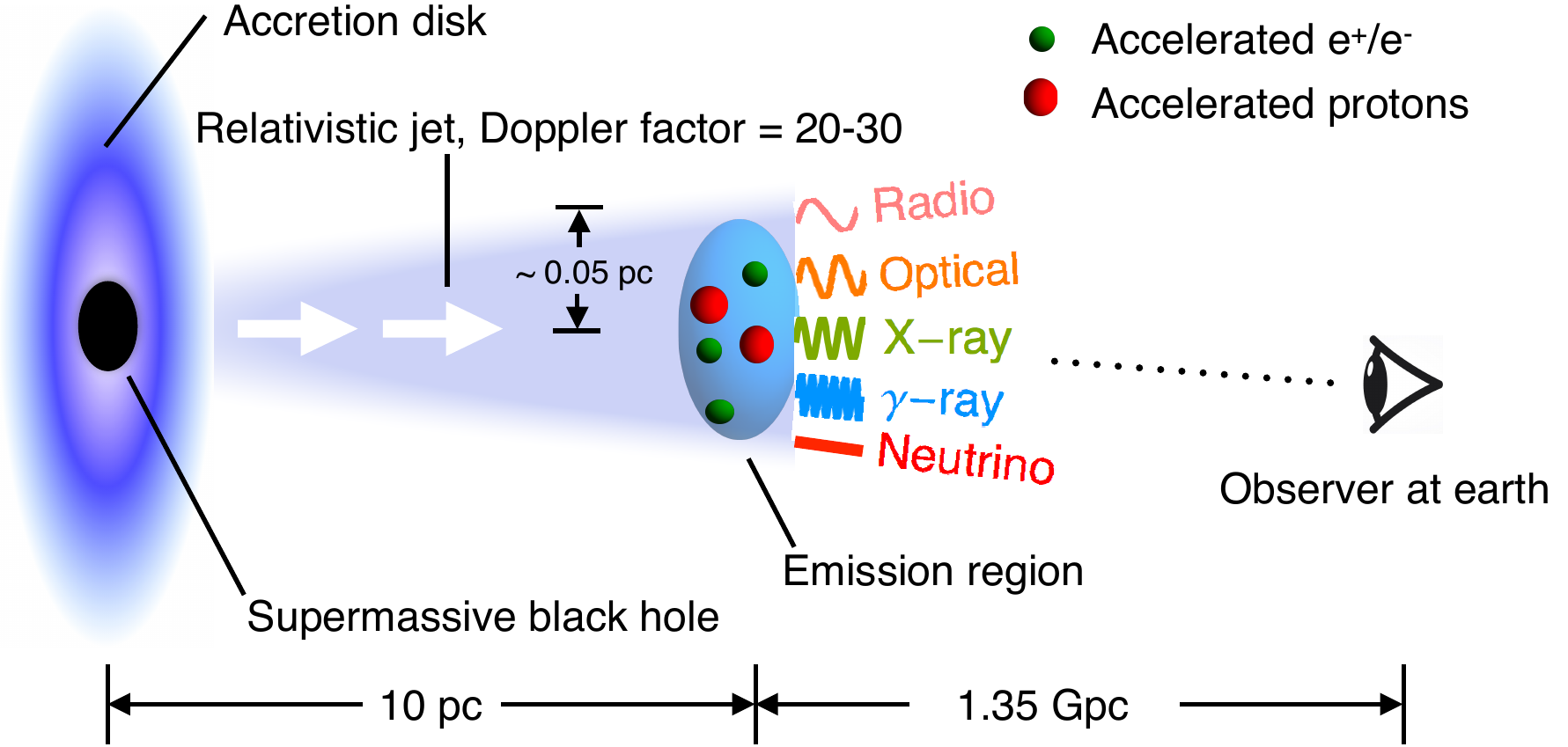}
\end{center}
\caption{\label{fig:illustration} {\bf Illustration of the emission region of TXS0506+056 traveling at relativistic speed. \rm The distance between radiation zone and the central black hole is not an explicit model parameter and given here only for illustration. Note that the physical sizes of various objects are not drawn to scale. }
}
\end{figure}

\begin{figure}[t]
\begin{center}
\includegraphics[width=0.45\textwidth]{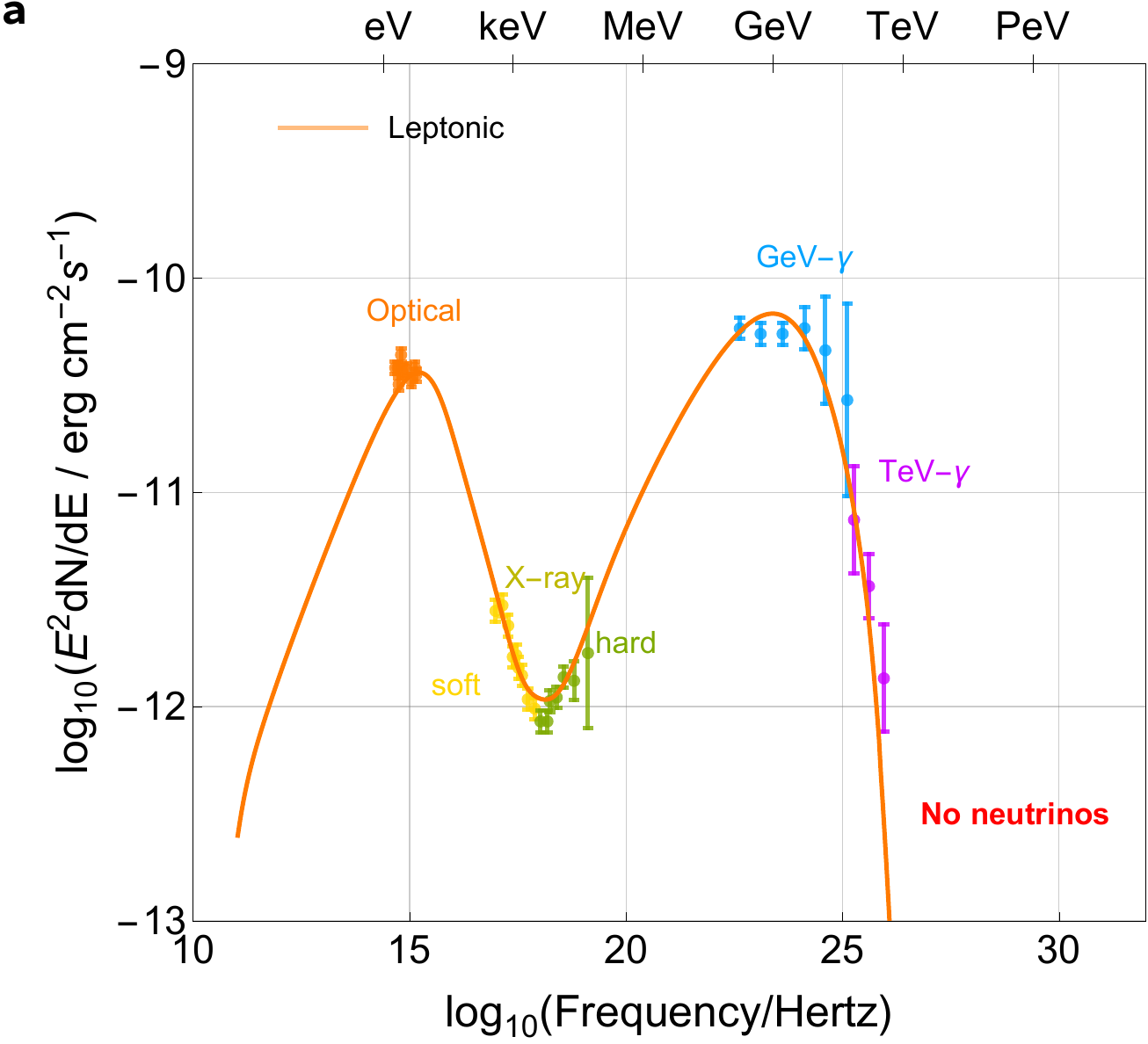}
\includegraphics[width=0.45\textwidth]{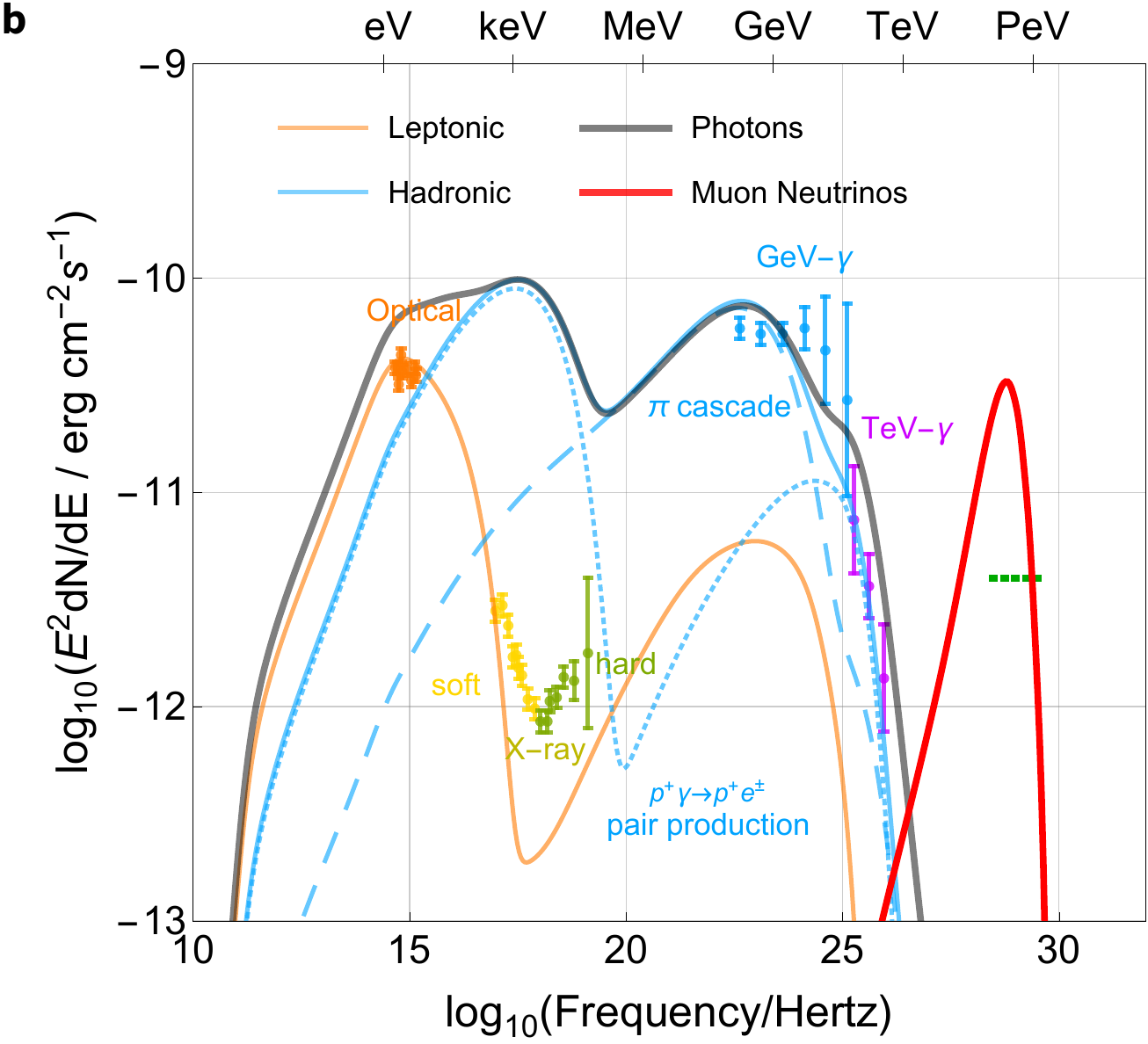}
\end{center}

\caption{\label{fig:sedpuremodels} {\bf Spectral energy flux from TXS0506+056 flare for two hypothetical scenarios.} The energy spectrum is well reproduced by a purely leptonic model (a) without neutrino production, whereas a simple hadronic model, in which the second hump comes from $\pi^0$ and $\pi^{\pm}$ decays, overshoots the observed X-ray flux (b). Data points reflect the observed flux and spectrum during the flare \cite{TXS_MM}. 
Colored curves indicate model components as given in the legend. The dashed horizontal green line corresponds to the expected level and energy range of the incident neutrino flux to produce one muon neutrino in IceCube in 180 days.}
\end{figure}

The amplification of radiation from the relativistic jet spawned by the central supermassive black hole in an active galactic nucleus makes the jet the dominant source of emission, if the observer has a frontal view of it, as is the case in a blazar like TXS0506+056. It is therefore appropriate to place the locale of particle acceleration and neutrino emission in the jet of TXS0506+056.  An illustration of the structure of a blazar and the location of the emission region is presented in \figu{illustration}.

The emission of very-high-energy radiation requires that the radiating particles, electrons or cosmic nuclei, be accelerated to even higher energies. A widely favored acceleration process is Fermi (diffusive) shock acceleration: charged particles gain energy by the frequent and repeated crossing of a shock front, leading to a particle spectrum in the form of a power law ($\propto E^{-\alpha}$) with $\alpha>1$; similar spectra are indeed observed in nature. Once accelerated, the energetic particles interact and radiate in a region referred to as the {\em radiation zone}. 

We shall now define physical scenarios of blazars that can be adapted to the observed properties of TXS0506+056. The first questions is whether or not they can reproduced these properties. If they can, the second question is where in the spectrum signatures of cosmic rays arise and what the source properties must be, given the observational constraints. The spectrum of electromagnetic radiation from AGN blazars has two characteristic components, a low-energy one arising from synchrotron radiation of energetic electrons, and a high-energy one typically attributed to Compton up-scattering of ambient photons by the same electrons (inverse Compton scattering) \cite{1995PASP..107..803U,2013ApJ...768...54B}; see \figu{sedpuremodels}a for a pictorial example; technical details can be found in the Methods section. Models of this type are collectively referred to as {\em leptonic} and are widely used to model the spectra of electromagnetic radiation from blazars.

The neutrino emission requires a {\em hadronic} scenario instead. Cosmic-ray nucleons
at energies $\sim10$~PeV will interact with UV photons to produce charged and neutral pions~\cite{Mucke:1999yb,Hummer:2010vx}. A charged pion decays (via a muon) into an electron or positron, which radiates just like any other electron, and three neutrinos that can travel to Earth and are a smoking-gun signature for the acceleration of cosmic nuclei. A neutral pion decays into two photons with similar energy as that of the neutrinos, providing a direct relation between neutrino and photon emission.
It is occasionally assumed that hadronic photon emission is responsible for the high-energy component of the spectrum~\cite{2016NatPh..12..807K}, inspired by the case of Mrk 421 which has a different SED that indeed allows this possibility, but a self-consistent analysis of all relevant processes indicates that the synchrotron X-ray emission by secondary electrons would unavoidably overshoot the observed flux~\cite{2017ApJ...843..109G,2015MNRAS.448.2412P}. For this reason we find that the flare state of TXS0506+056 cannot entirely be reproduced with a hadronic model, see \figu{sedpuremodels}b; an in-depth investigation on hadronic models is available in the {\em Supplementary Information}, see Supplementary Figures~3 and~4. This leaves the question what the maximal neutrino flux during the flare can be, and what the photon signature of a hadronic model actually is. The same constraint applies to the quiescent state, although it is weaker there. Instead, both the quiescent and the flare state are easily described by a leptonic scenario (see \figu{sedpuremodels}a for an example). 

\begin{figure}[t]
\begin{center}
\includegraphics[width=.6\textwidth]{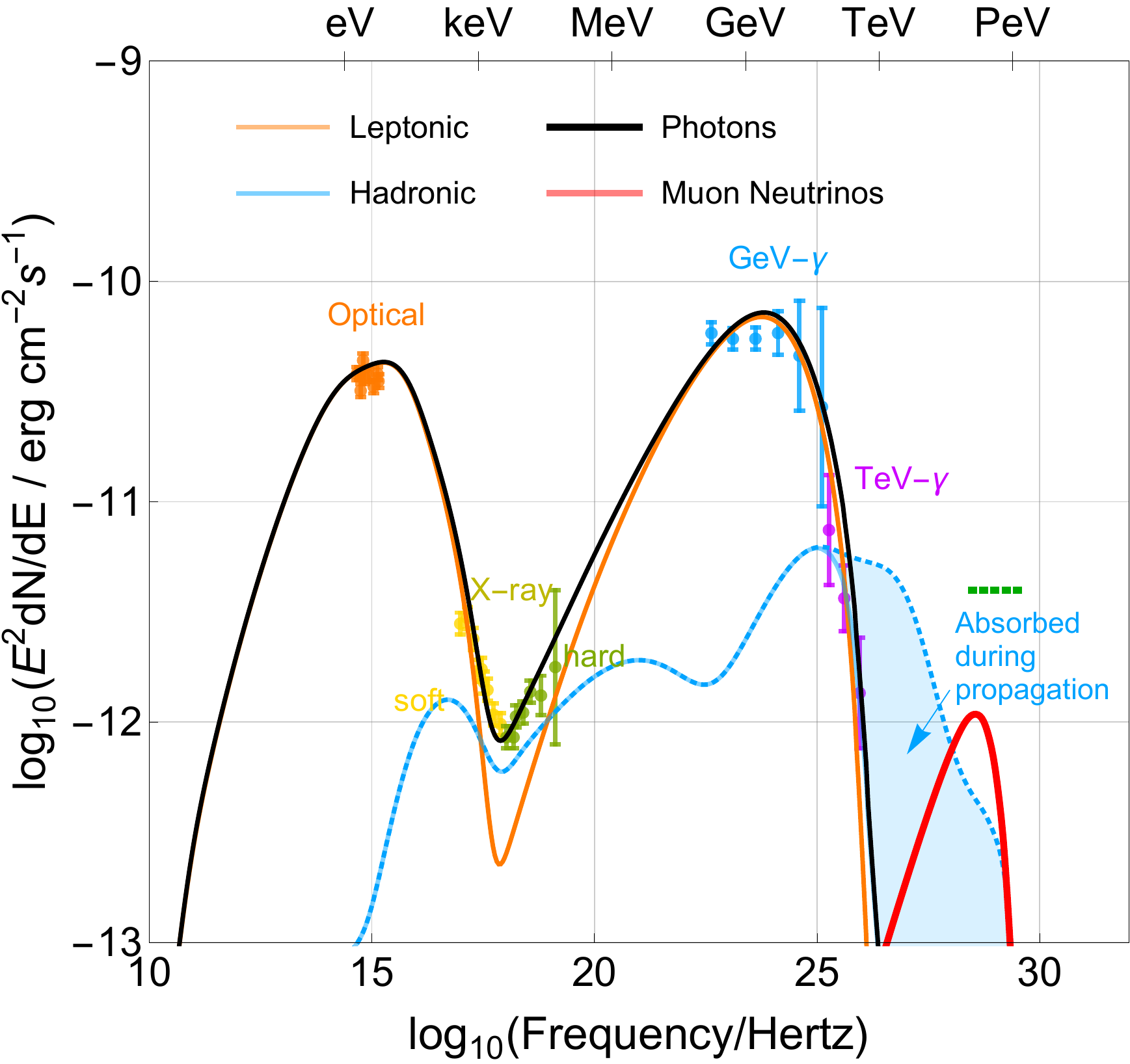}
\end{center}
\caption{\label{fig:hybridfit} {\bf Energy flux from TXS0506+056 across the electromagnetic spectrum and for neutrinos.} 
Here the energy spectrum is modeled in our hybrid scenario with both leptonic and hadronic contributions. High-energy photons are absorbed during propagation by extragalactic background light, here indicated by the blue shaded region and modeled as in \cite{2013ApJ...768..197I}.  Data points reflect the observed flux and spectrum during the flare \cite{TXS_MM}. 
The dashed horizontal green line corresponds to the expected level and energy range of the incident neutrino flux to produce one muon neutrino in IceCube in 180 days.
}
\end{figure}

We propose the {\em hybrid model} displayed in \figu{hybridfit}, in which the bulk of photon emission is of leptonic origin, and hadronic contributions are as strong as permitted by the X-ray data. In addition, the outflow of matter and radiation in the blazar should not be so powerful that the surrounding matter is blown away, otherwise the activity of the central black hole would be quenched, which for continuous and isotropic emission leads to the so-called Eddington limit. Modeling the flare on the basis of an increase in the particle-acceleration power alone will invariably require a jet power that is in excess of the Eddington luminosity by several orders of magnitude as discussed in the {\em Supplementary Information}, see Supplementary Figure~1. It is known that so-called proton synchrotron models may alleviate the tension with the Eddington limit \cite{2015MNRAS.448..910C}, but they do not simultaneously reproduce the SED of TXS0506+056 and the energy and flux of neutrino emission, see Supplementary Figure~4; similar arguments apply to models with higher maximum proton energy, see Supplementary Figure~2. During outbursts or for a collimated outflow in a jet the Eddington luminosity may be exceeded, because in the former situation the stored energy can still be radiated away and in the latter case the jet does not interfere with the accretion flow. The power excess is probably moderate and within a factor of ten (see, e.g., \cite{2015MNRAS.453.3213S}). A model that satisfies the observational and the power constraints requires the flare to be produced by an increase of electron and proton injection power in a smaller core of the radiation zone. We also allow for an increase of the magnetic-field strength, as that usually goes hand-in-hand with enhanced particle acceleration \cite{2018MNRAS.473.3394V}. 

Whereas the hadronic contribution to the energy spectrum is clearly visible in the X-ray band, it is also present in the TeV band, but here attenated by pair production with the extragalactic background light. Depending on the shape of the SED and the distance to the source the hadronic TeV-scale emission can be sizable, for example for nearby radio galaxies \cite{2017AIPC.1792e0027C}. The neutrino emission (red curve) is below that corresponding to one observed neutrino above 100~TeV in 180 days, and it is limited by the observed X-ray emission on account of a correlation between the neutrino response and that in the X-ray band. The GeV-band gamma-ray flare arises from enhanced electron injection. 
Increasing the proton production only would result in ``orphan'' neutrino flares coinciding with intense X-ray emission and, likewise, an increase in the electron acceleration power might account for the known orphan GeV/TeV gamma-ray flares \cite{2004ApJ...601..151K}. 

The multi-TeV photon emission is of particular interest in view of a puzzling finding. TeV gamma-ray emission from blazars is partially absorbed in the intergalactic medium on account of interactions with ambient optical and infrared radiation. After correction for this absorption, the spectra of blazars suggest the existence an additional radiation component in addition to that expected with leptonic models \cite{2014ApJ...785L..16A}, which may indicate for the existence of new elementary particles such as axions \cite{2009PhRvD..79l3511S} or blazars emitting a very powerful stream of cosmic rays at the highest energies \cite{2010APh....33...81E}. We posit that the hadronic interactions leading to neutrino emission around 200~TeV offer a simpler and more natural explanation, that predicts time variability in line with that observed.

\begin{figure}[t]
\begin{center}
    \includegraphics[width=.5\columnwidth]{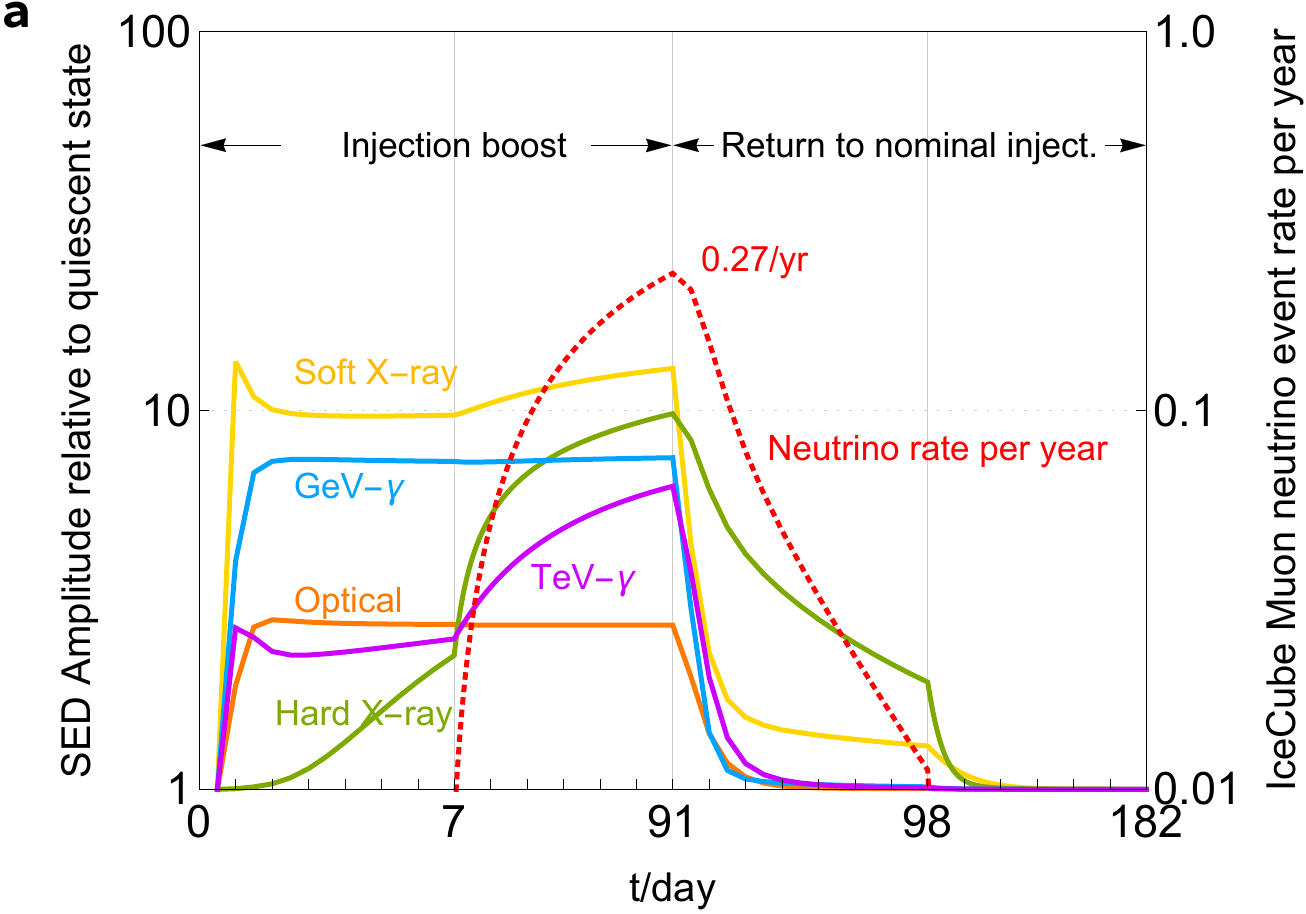}
    \hspace{0.5cm}
	\includegraphics[width=.4\columnwidth]{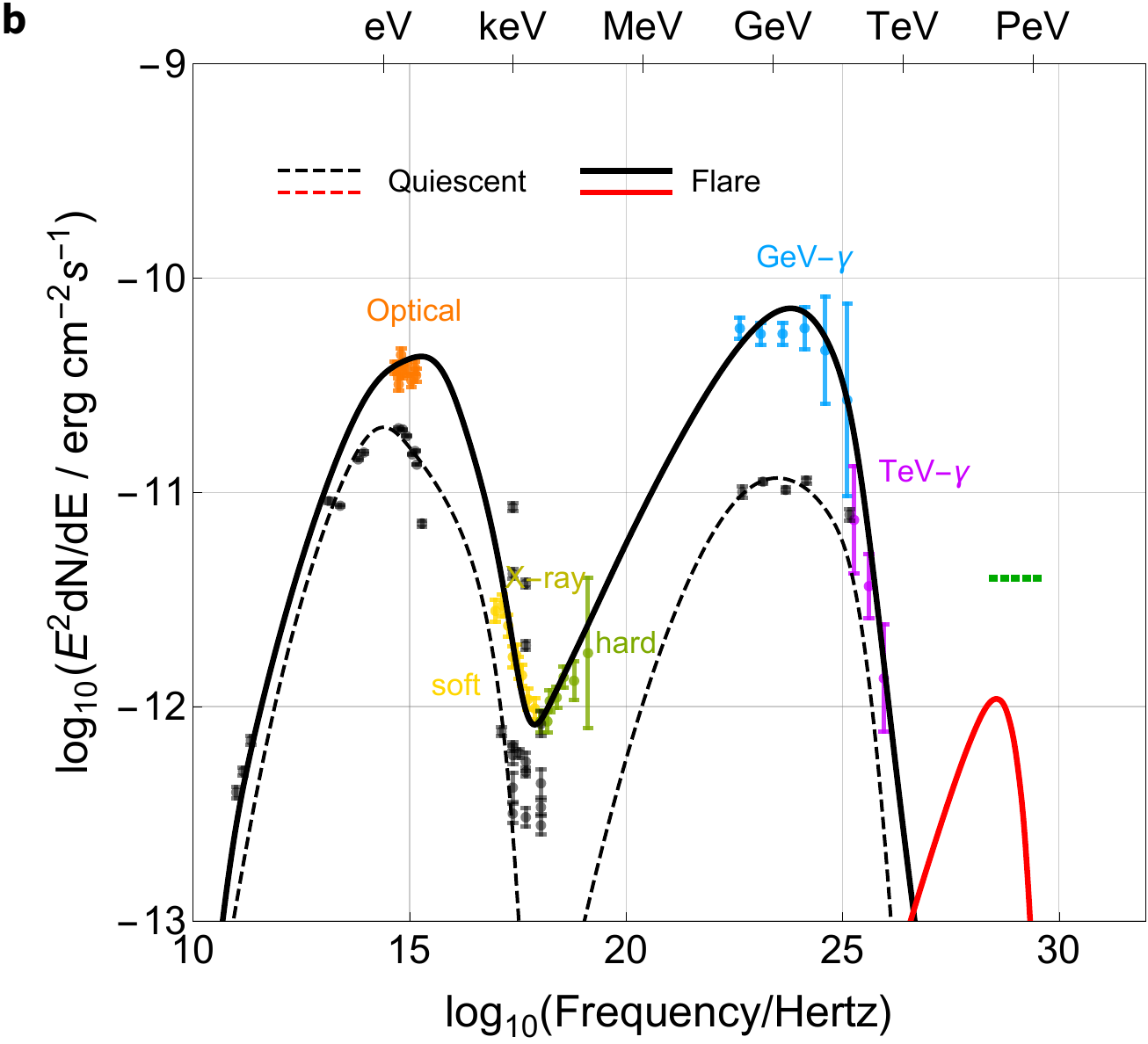}
\end{center}
	\caption{\label{fig:temp_behavior} {\bf Time-dependent simulation of the lightcurve during the flare.} The response is shown for an example period of 90 days. a) Temporal response in 250-TeV neutrinos and various wavelength bands. Note the scaling variations in the time axis. b) Spectral response of the signals in neutrinos and photons. The black dots (including the corresponding uncertainties) reflect data taken during the years prior to the flare.}
\end{figure}

\figu{temp_behavior}a displays the amplification of the signals in various wavebands and in neutrinos for an assumed flare duration of 90 days. Any short-term variations in the particle injection rate would affect the radiation flux with the same response time as shown in the figure, e.g., swiftly in soft X-rays, slower in hard X-rays, and slowest in neutrinos on account of the low energy-loss rate of protons. To be noted are the strong enhancement in the neutrino flux and the flux correlation between neutrinos, hard X-rays, and TeV gamma rays. The neutrinos are produced in interactions with hard X-ray photons and hence their flux receives a synergistic boost due to the increased densities of both, the target photons and the protons. In the case of leptonic emission, some of the gain is lost on account of enhanced energy losses. After the additional injection into the core vanishes, the electrons rapidly cool and consequently the target photon density for the remaining cosmic rays decreases to the quiescent level. The neutrino emission continues at low rate in the larger radiation zone.   
\figu{temp_behavior}b shows the spectral energy distribution in photons and neutrinos before the flare, at its peak, and late in the cool-off phase. The steeply falling spectrum around the threshold energy of the TeV gamma-ray and soft X-ray telescopes implies that small variations in the injection can lead to sporadic changes in the measured signal, in line with those observed.

The time dependence of the SED of TXS0506+056 supports our focus on so-called Synchrotron\allowbreak -Self-Compton (SSC) models, in which the high-energy photon peak is the result of up-scattering of low-energy synchrotron photons that originate from the same electron population. In this case and for the Thomson scattering that applies here, the enhancement of the second peak is expected to scale with the square of the increase of the synchrotron photons, as is observed.

Our model allows for 0.27 muon neutrinos per year for energies $E_{\nu}>120$ TeV during the flare state, which lasted at least half a year, given the GeV-band lightcurve, and so we expect more than 0.14 neutrino events in that period, implying a probability higher than 10\% to actually detect a neutrino. The model describes the dramatic enhancement of the neutrino flux during the entire duration of the flare, and renders the two-week delay between the neutrino detection and the TeV-band activity seen with MAGIC \cite{TXS_MM} insignificant, likewise that with the later detection with VERITAS \cite{2018ApJ...861L..20A}. The late detection in both bands after a few months of GeV-band flaring reflects the slow response predicted by our model. In any case, the time-dependent analysis shows that variability in the band 100~GeV to 1~TeV arises from both leptonic and hadronic contributions. The acceleration of nuclei during quiescence could result in weak neutrino emission, suggesting that other neutrinos from TXS0506+056 might be hidden in the IceCube data, despite the non-detection of other blazars \cite{2017ApJ...835..151A}.

We demonstrated that the coincidence of a neutrino with a flare from TXS0506+056 can be described by a significant increase of the injection rates of cosmic nuclei and electrons. This provides evidence for the acceleration of cosmic rays up to energies of about 10 PeV in certain AGN flares, and there is no evidence for a connection to ultra-high energy cosmic rays at energies above 100~PeV (the consequences of this scenario are outlined in the {\em Supplementary Material}, see Supplementary Figure~2). Efficient neutrino production requires either a more compact production region during the flare, such as the denser core of a larger radiation zone, or an injected proton luminosity far in excess of the so-called Eddington limit. Since the production of neutrinos necessarily implies the emission of high-energy photons, electrons and positrons, the resulting electromagnetic cascades must be visible as X-rays and also TeV gamma-rays -- thus constraining the maximally allowed contribution of photo-hadronic interactions and consequently the expected neutrino flux. When taking all constraints into account, we find predicted neutrino rates significantly lower than, but still statistically consistent with one event per year. Our preferred model describes how the neutrino flux, and to a lesser degree also that of hard X-rays, is over-proportionally enhanced during the flare, if that is sufficiently long-lived, explaining why neutrinos are found during such flares and are otherwise statistically not attributable to blazars. Our time-dependent modeling of the relevant physics processes provides a self-consistent picture for TXS0506+056 that is based on observations of neutrinos and photons in all spectral bands.

\subsection*{Correspondence} 

Correspondence about the manuscript should be directed to S.G. or A.F. 

\subsection*{Acknowledgements} 

 We would like to thank Andrew Taylor, Andrea Palladino and Elisa Bernardini for useful discussions and comments. 
 S.G., A.F. and W.W. have received funding from the European Research Council (ERC) under the European Union’s Horizon 2020 research and innovation programme (Grant No. 646623).

\subsection*{Author contributions}

S.G. performed the numerical modeling and artwork. A.F. extracted and analyzed the data. S.G. and A.F. provided first technical documentation. All authors contributed to the development of the theoretical ideas and the interpretation of the results. The text of the final manuscript was written by M.P., W.W. and A.F. with contributions from S.G.

\subsection*{Competing interests}

The authors declare no competing interests.

\section*{Methods}

The time-dependent radiation modeling is performed with the numerical code {\sc AM}$^3$ (Astrophysical Multi-Messenger Modeling) that has been applied to a similar physical environment and is documented in \cite{2017ApJ...843..109G}. This Methods section goes into detail on the construction of the spectrum from a publicly accessible data stream called {\it The Astronomical Telegram} and the derivation of model parameters.

\subsection*{Construction of the observed energy spectrum}

Initial information about the IceCube event 170922A has been shared on {\it The Gamma Coordinates Network} (GCN) notice 21916 \cite{IceCube_GCN}, providing the time, direction, and angular uncertainty of a muon track, a secondary product of a neutrino interacting in the rock and the ice around the detector. The muon track deposited $23.7 \pm 2.8$ TeV of energy in Cherenkov light, meaning that the true neutrino energy is higher, most likely around $290$ TeV and with a small probability above 1 PeV. The purpose of the GCN public data stream is to alert the astronomical community about potential {\it targets of opportunity}, astrophysical events that might be worth studying across multiple wavebands. The present neutrino event triggered immediate follow-up observations. A six-fold increase, compared to the 3FGL catalog value, in the 0.1 -- 300 GeV flux from the blazar TXS0506+056 located inside the directional error circle of the neutrino event was reported, and GeV-band flaring had been ongoing for a few months \cite{2017ApJ...846...34A} . A key ingredient for building our model is the joint Swift XRT and NuSTAR observation of the soft and hard X-ray emission. Two weeks after the initial alert the MAGIC telescope detected gamma-rays above 100 GeV with a very soft spectrum, whereas the source had been invisible in TeV gamma-rays before \cite{TXS_MM}.

From the visual inspection of the light curves, we conclude that the brightening of the spectrum started in June 2018, reaching peak luminosities close to the date of the neutrino detection, slowly decreasing thereafter without returning to the previously observed quiescent levels. The continuous activity is confirmed by later observations \cite{2017ApJ...846...34A,2018ApJ...861L..20A}. The light curves are not smooth and show high stochastic variability from one week to the next. 

For the temporal evolution study in \figu{temp_behavior}, we construct the spectrum of the quiescent state from archival observations available from the database of the Space Science Data Center (\href{http://tools.asdc.asi.it/}{SSDC}) and from NASA/IPAC Extragalactic Database (\href{https://ned.ipac.caltech.edu/}{NED}). These data are not strictly contemporaneous and may be partially contaminated by previous flares.

Model parameters are obtained by $\chi^2$ minimization for the spectral data. To reduce the bias arising from different sparsity of data across the electromagnetic spectrum, each characteristic waveband is represented by a bow-tie, approximate power-law bands corresponding to the integrated flux and the spectral index with their uncertainties. The fit minimizes the integrated flux and the average spectral index of the SED to the power-law fits obtained by the experiments in each spectral band. Radio data are taken as upper limit, since the radio emission typically arises a much larger region than the radiation zone \cite{2016NatPh..12..807K}. While this method successfully constrains the parameters, it is based on a simplified representation of the data and does not account for systematic uncertainties. The consequences for the interpretation of the $\chi^2$ values of our analysis are discussed in the \emph{Supplementary Information}.

\subsection*{Model for the emission zone}

The emission from AGN blazars is dominated by that produced in their jets, on account of the strong relativistic Doppler amplification. The observation of rapid time variability in the observed radiation flux implies a very compact emission region that is tiny compared with the jet. As the escape of radiation from this emission region is typically much faster than acceleration-rate changes and the energy loss of cosmic rays, one neglects the internal structure and models the emission zone as a spherically symmetric, homogeneous blob of radius $R_\mathrm{blob}^\prime$ that is filled with gas, photons, magnetic field, and energetic particles. The entire emission zone moves with Doppler factor $\Gamma_\mathrm{bulk}$, and we denote with primes ($^\prime$) physical variables in the jet rest frame. Electrons and ions are continuously and isotropically injected with power (or luminosity) $L_\mathrm{inj}^\prime$, and their injection rate obeys a power law, $d^2n^\prime/d\gamma^\prime dt^\prime=K^\prime\,{\gamma^\prime}^{-\alpha}$, where $K^\prime$ is a function of the injection power and $\gamma^\prime$ is the Lorentz factor of the particles that is allowed to take values in the range $\gamma_{\mathrm{min}}^\prime$ and $\gamma_{\mathrm{max}}^\prime$. 
The emission region is assumed to be filled with a homogeneous, randomly oriented magnetic field of strength $B^\prime$. Energetic particles are allowed to leak out on the time scale $t_{\mathrm{esc}}^\prime = R_\mathrm{blob}^\prime/(\eta_\mathrm{esc}c)$, where $\eta_\mathrm{esc}\le 1$ is treated as a free parameter.

We model all relevant interactions of the particles, which for electrons include synchrotron emission and absorption, inverse-Compton scattering, and pair production and annihilation. For protons we account for Bethe-Heitler pair production ($p + \gamma \longrightarrow p + e^\pm$) and photo-pion production ($p + \gamma \longrightarrow p + \pi$). The pions decay to eventually yield neutrinos, electrons, and positrons.

For the hybrid model we model the quiescent state by injecting electrons and protons into the blob until an equilibrium between injection, cooling and particle escape is reached. The injection power of protons is limited by the Eddington luminosity in this phase. The model SED is then fitted to that observed. The flare is initiated in a smaller core in the radiation zone that characterizes the quiescent state, and both contribute to the observed emission in a superimposed way, a scenario that mimics localized particle acceleration and subsequent diffusive transport into a larger emission zone \cite{2015MNRAS.447..530C,2016MNRAS.458.3260C}.
The radiation from the core is significantly brighter, rendering the presence of the larger (quiescent) zone insignificant. It can be shown that for typical escape parameters the radiation of protons leaking from the core into the outer blob is negligible because of the much smaller radiation density. The transport of electrons from the core into the outer zone contributes less than 16\% of the nominal quiescent-state radiation.
We increase the injection of electrons by a factor of 3, that of protons by a factor of 10, and the magnetic-field strength by a factor of 20 (see \eTab~1). Enhanced activity in the smaller core persists for a certain period of time (here $t=90$ days, which is our conservative example is much less than the total duration of the flare. The stochastic variability seen in the light curves during an extended flaring period may be the result of the fluctuations in the acceleration rate. During the enhanced activity in the core the model allows the total jet power to exceed the Eddington luminosity. This picture is supported by the Fermi GeV-band light curve, which shows that the peak brightness is reached one week before the neutrino event, decreasing after a few days. The enhanced emission and variability continues for several more months \cite{2018ApJ...861L..20A}.

\subsection*{Determination of model parameters}

Most parameters of the hybrid model are obtained through extensive parameter scans using the time-dependent {\sc am}$^3$ code. We use the previously described $\chi^2$ optimisation to determine the goodness of fit for a particular SED. The blob size, the Doppler factor, the effective escape velocity, $\eta_\mathrm{esc}$, and the maximal proton energy are then adjusted with a view to maximize the neutrino flux in the relevant energy range ($>120$ TeV) and to minimize the required jet power.

Hybrid simulations require the primary electron spectrum to follow a broken power-law: $dN^{\prime}/d\gamma_{e}^{\prime}\propto \gamma_{e}^{\prime \alpha_{e} }$ where $\alpha_{e}=\alpha_{e,1}$ for $\gamma_{e,\mathrm{min}}^{\prime}<\gamma_{e}^{\prime}<\gamma_{e,\mathrm{br}}^{\prime}$ and $\alpha_{e}=\alpha_{e,2}$ for $\gamma_{e,\mathrm{br}}^{\prime}<\gamma_{e}^{\prime}<\gamma_{e,\mathrm{max}}^{\prime}$. For the leptonic or hadronic model, a single-power-law injection of both electrons and protons describes the relevant parts of the spectrum sufficiently well. The determination of the parameter values in \eTab~1 requires deterministic scans of the higher-dimensional parameter space, for which we performed $\mathcal{O}(10^8)$ individual {\sc am$^3$} simulations.

For TXS0506+056, we find a good fit for a Doppler-factor {$\Gamma_\mathrm{bulk}=28.0$}. The redshift of the host galaxy is known as $z=0.34$ \cite{2018arXiv180201939P}. With the typical simplification that the highest contribution to the neutrino flux originates near the pion-production threshold ($\Delta$-resonance approximation), the proton energy in the observer frame is $E_p^{\rm obs} \approx 20 E_\nu^{\rm obs}$. The maximal gamma factor of protons in the comoving frame is $\gamma^{\prime}_p \simeq (E_p^{\rm obs}/m_p)(1+z)/\Gamma_\mathrm{bulk} \approx 2 \cdot 10^5$. The typical energy of the target photons in $p\gamma$ interactions is $\epsilon_{\gamma,{\rm target}}^\prime \simeq E_{\rm threshold} / (2 \gamma_p^\prime) \approx 2.5\cdot 10^{-6}\,(0.2\, {\rm GeV}) = 0.5\, {\rm keV}$. In the observer frame this corresponds to $\epsilon_{\gamma,{\rm target}}^{\rm obs} \simeq \Gamma_\mathrm{bulk} \epsilon_{\gamma,{\rm target}}^\prime / (1 + z) \approx 15\, {\rm keV}$, i.e. hard X-rays. Our full simulation is based on the realistic $p\gamma$ cross section and multi-pion emission at higher energies, yielding for the target photons a wide range of energies in the hard X-ray band.
A higher maximum energy of protons leads to a higher peak energy of the neutrino spectrum (even after a possible re-normalization of the injection spectrum), which is incompatible with the current neutrino data as we discuss in Supplementary Figure~2, and so there is no direct relation between the neutrino event in question and the origin of ultra-high-energy cosmic rays.

The parameters obtained for the different models are listed in Suppplementary Table~1.

\subsection*{Computation of neutrino rates}

For the computation of the expected neutrino event rate in IceCube we use the effective area reported in \cite{2017arXiv171206277I}. It is the highest effective area published by IceCube and valid for transient astrophysical sources. The expected atmospheric neutrino background rate is estimated with the numerical code {\sc MCEq} \cite{Fedynitch:2015zma}, the GSF cosmic-ray flux \cite{Dembinski:2017zsh}, and the {\sc SIBYLL-2.3c} hadronic interaction model \cite{Riehn:2017mfm}; it lies in the range $0.003$ -- $0.001$ neutrino tracks per year depending on the assumed minimal energy between $180$ -- $300$ TeV for a solid angle of $0.97$ square degrees \cite{TXS_MM}, commensurate with the 90-\% directional uncertainty of the IceCube event. Using a probability distribution of true neutrino energies based on \cite{2018arXiv180107277P}, the background rate is 0.001 events per year. The hybrid radiation model predicts a neutrino-event rate at $E_{\nu}>180$~TeV of $1.4\times10^{-4}$ tracks per year during quiescence and a peak rate of $0.27$ tracks per year for the recent flare. By using a signal-over-background definition, the significance for this particular neutrino to originate from the TXS0506+056 flare reaches the $2.8\, \sigma$ level for a true neutrino energy $E_\nu > 180$ GeV.

\subsection*{Data availability}

The historical observations analyzed during the current study are available in the 
\href{https://tools.asdc.asi.it/SED/sed.jsp?ra=05+09+28.4&dec=%2B05+42+01.0}{SED Builder Tool of the Space Science Data Center (SSDC)} and from the \href{https://ned.ipac.caltech.edu/cgi-bin/datasearch?search_type=Photo_id&objid=177466&objname=WISE%20J050925.96%2B054135.3&img_stamp=YES&hconst=73.0&omegam=0.27&omegav=0.73&corr_z=1&of=table}{NASA/IPAC Extragalactic Database (NED)}. 

The data that support the plots within this paper and other findings of this study are available from the S.G. and A.F. upon reasonable request.

\renewcommand{\figurename}{Supplementary Figure}
\renewcommand{\tablename}{Supplementary Table}

\begin{appendix}

\section*{Supplementary Information} 

The sparsity and uncertainties of the TXS0506+056 observations will necessarily allow for some model-parameter variations and potentially for degenerate interpretations, and it is prudent to check whether other possible interpretations of the observed photon flux and neutrino emission might exist. In this section we discuss a set of scenarios in the light of a correlated neutrino flux.

Concerning the observation, one source of uncertainty can be attributed to the instrumental precision and the data analysis, and another part is related to the non-simultaneity of the measurements. The first type of uncertainty is taken into account as a penalty in the $\chi^2$ minimisation. While a rigorous treatment of the second type of uncertainty is beyond the scope of this work, it is possible to study this aspect using semi-analytical methods. 

Concerning the modeling, the large parameter space naturally allows for some degeneracy in the description of the SED. The common scenarios include (1) the favored hybrid model with a limited proton maximal energy, $ E_{\rm p, max}$, (2) a hybrid model using a single radiation zone without core, (3) a leptonic (SSC) model, (4) a hybrid model with $E_{\rm p, max}$ similar to that of ultra-high-energy cosmic rays (UHECR, about $10$~EeV), (5) a fully hadronic model, and (6) a proton-synchrotron scenario. 

The preferred hybrid scenario with an imposed limit on $E_{\rm p, max}$ (1) is comprehensively described in the main text and the \emph{Methods} section. The leptonic (SSC) model (3) can reproduce the SED but is not explicitly discussed, since there is no neutrino emission. In the next subsections we test the viability of the remaining four scenarios.

\subsubsection*{Hybrid one-zone model}
\begin{figure}[t]
    \begin{center}
        \includegraphics[width=.53\columnwidth]{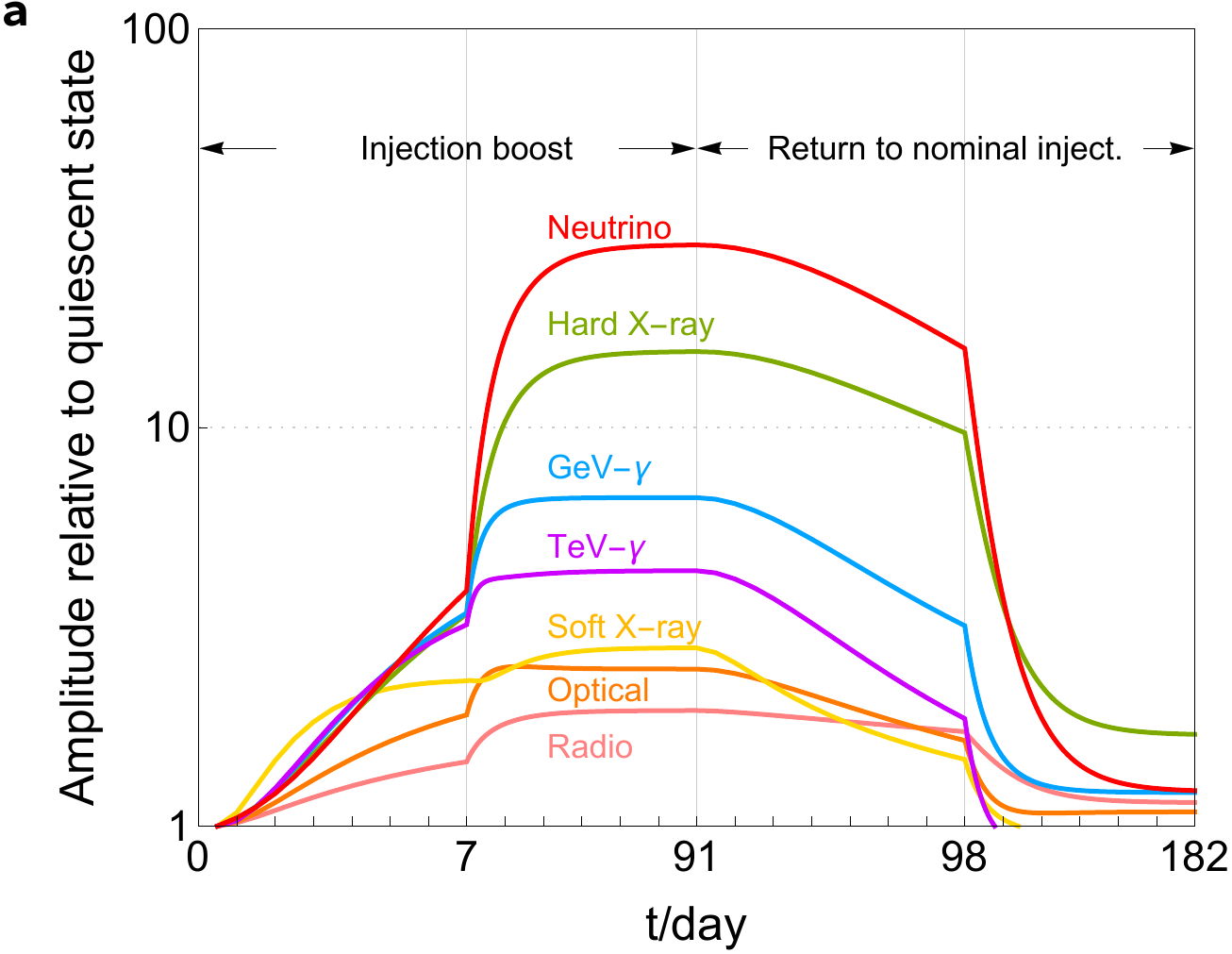}
        \includegraphics[width=.463\columnwidth]{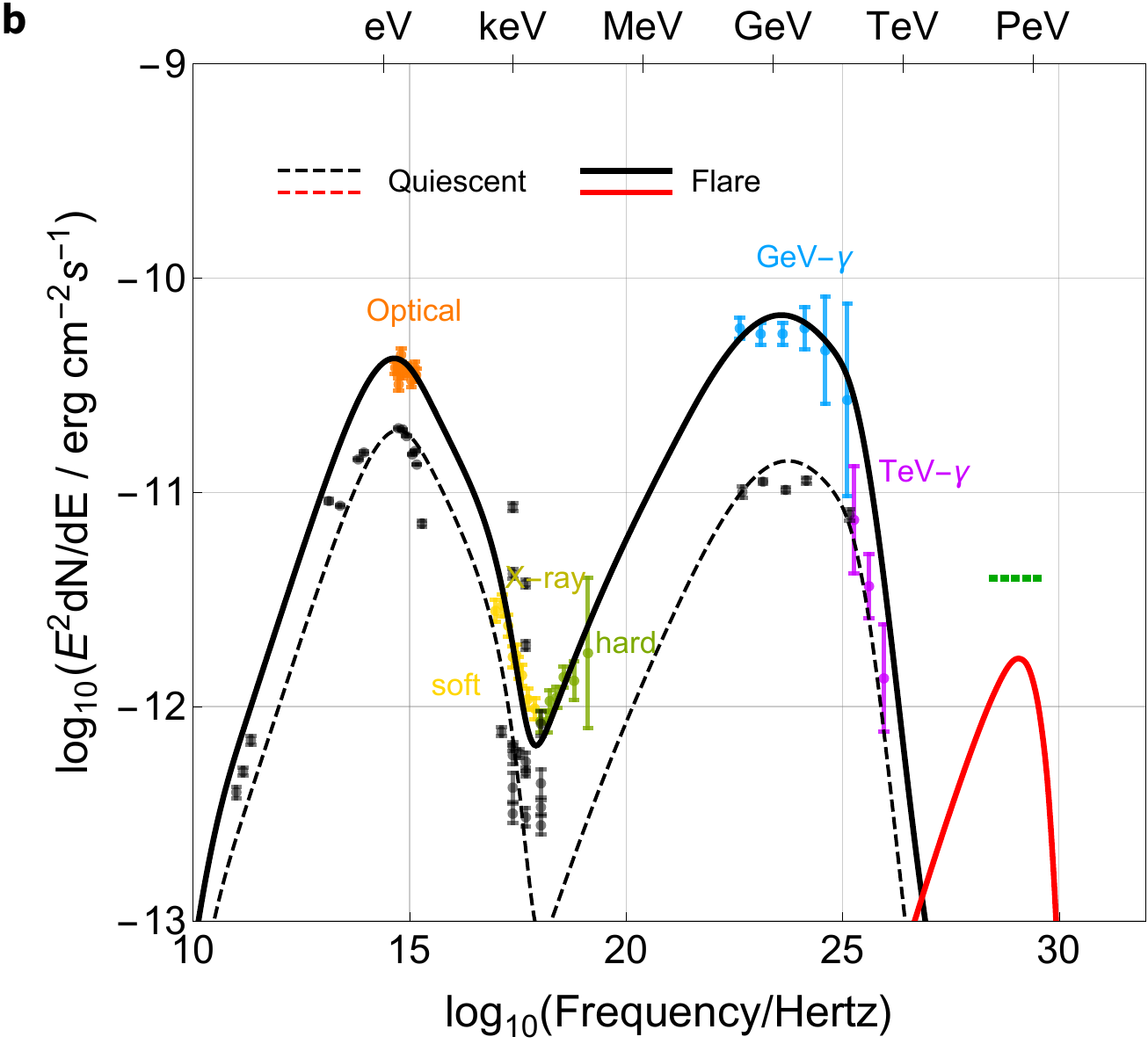}
    \end{center}
        \caption{\label{fig:temp_behavior_onezone} {\bf Time-dependent simulation of the flare with a single radiation zone}. As in Fig.\ 4 of the main text, a) displays the temporal response and b) shows the spectral response. This model elegantly explains the transition between quiescent and flare SEDs of through a simultaneous increase of injection power of protons and electrons by a factor of three. However, the required power to reproduce the flaring SED and the neutrino observation would strongly exceed the Eddington luminosity.}
\end{figure}

The SEDs for the quiescent phase and the flare can be reproduced with a single radiation zone, simply by increasing the power of particle injection into the radiation zone which leads to the temporal responses in \efigu{temp_behavior_onezone}a. The expected neutrino flux is higher than in our baseline model and more closely matches that observed with IceCube and the SED can be nicely reproduced in quiescent and flaring states with similar sets of parameters, see \efigu{temp_behavior_onezone}b. As a major drawback, a large particle-injection luminosity is required that is far in excess of the Eddington luminosity for a black hole of $5\times10^{9}$ solar masses ($L_{p,\mathrm{inj}}=\Gamma_\mathrm{bulk}^2\,L_{p,\mathrm{inj}}^\prime = 10^{50.5}$ erg/s in the AGN frame \cite{2014Natur.515..376G}, compared to $L_\mathrm{Edd}=10^{47.8} $ erg/s). This interpretation therefore implies an accretion rate exceeding the Eddington luminosity by nearly three orders of magnitude at least during the flare, which is frequently considered to be unlikely for AGN blazars. Note, however, that such high excesses are obtained for cataclysmic sources, such as jets from tidal disruptions of massive stars~\cite{2018MNRAS.478.3016W}.

\subsubsection*{Hybrid model with UHECR interactions}

\begin{figure}[t]
	\centering 
    \includegraphics[height=7.5cm]{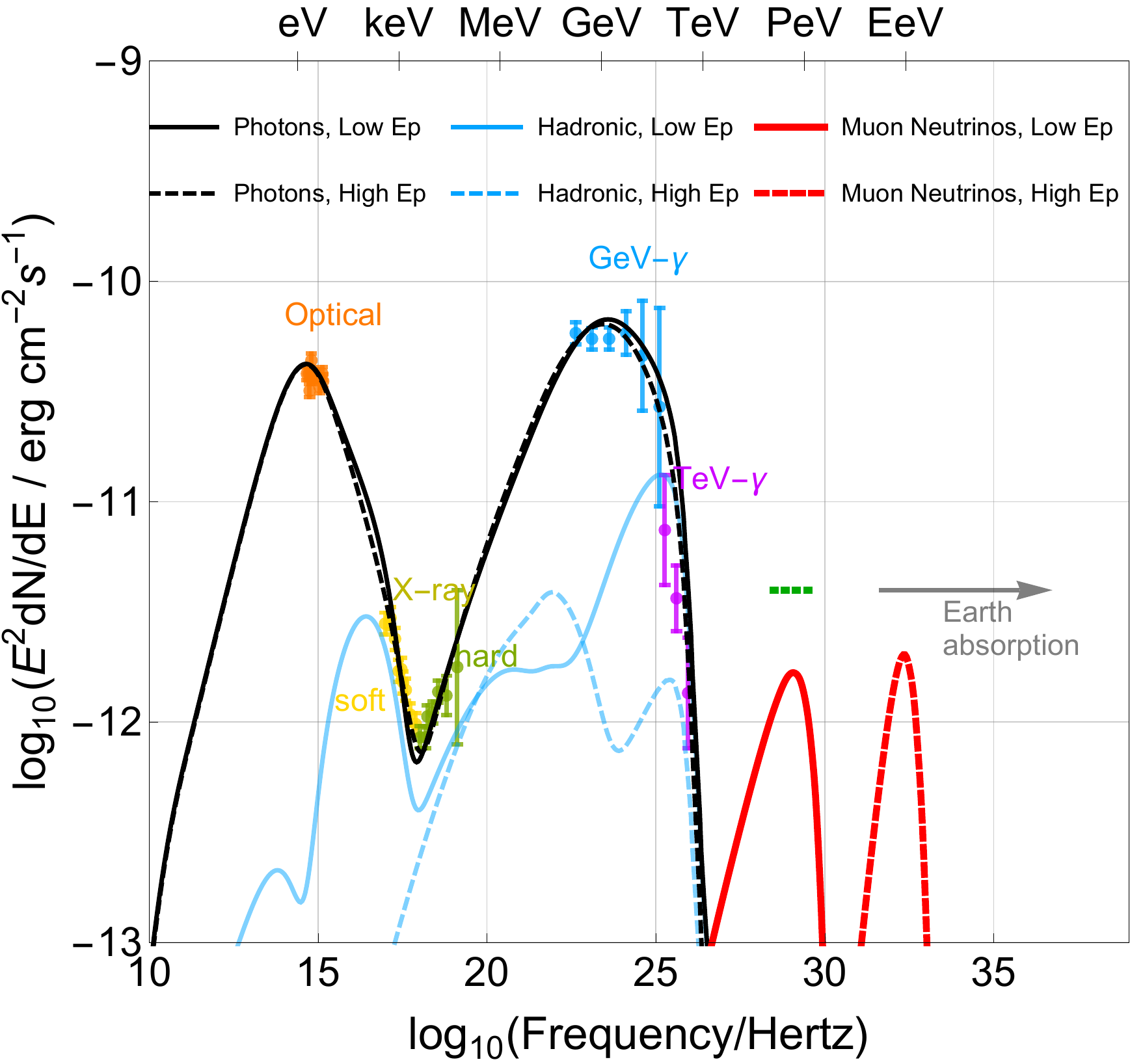}
	\caption{{\bf Hybrid model version including UHECR interactions in the source.} The solid curves refer to the hadronic components of the favored model (identical to Fig.\ 3) with the injection of protons up to $E_{\rm p, max} \sim 4.5$ PeV. The dashed lines show the impact of a proton population that extends up to UHECR energies $E_{\rm p, max} \sim 17$ EeV.
	\label{fig:UHECR}}
\end{figure}
The most relevant parameters related to neutrino production are the maximum proton energy, $E_{\rm p, max}$, and the proton injection luminosity, $L_{p,\mathrm{inj}}^{\prime}$. Variation of these parameters results in different spectra of hadronic photon and neutrino emission, but the photon SED may still be compatible with that observed. The choice of the maximal energy is restricted by the number and energy of the observed neutrinos and can be roughly approximated by $E_{\rm p, max} \approx 20~\langle E_\nu \rangle \simeq 4.5$ PeV. If the maximal energy is computed using the maximum acceleration rate and the light crossing time of the source (as commonly and highly optimistically assumed and known as the Hillas limit), protons may reach UHECR energies, $E_{p,\mathrm{max}}=\Gamma_\mathrm{bulk} \,(10$~EeV). \efigu{UHECR} demonstrates that in a hybrid scenario the photon SED would be well reproduced; the proton injection power would be low, $L_{p,\mathrm{inj}}^{\prime}=10^{43.9}~\mathrm{erg/s}$, otherwise the electromagnetic cascade emission would exceed the measured flux in hard X-rays and the TeV band. The scenario is not acceptable though, because the resulting neutrino flux peaks at a much higher energy, at a few EeV, an energy band in which all neutrinos are effectively blocked by earth along the path of propagation, since the source is located slightly below the horizon. For a deposited energy of 23.7$\pm$ 2.8 TeV of the muon track in IceCube, the incident neutrino energy $E_{\nu}$ is expected to lie between 183 TeV and 4.3 PeV at 90\% confidence level \cite{TXS_MM}. The expected event rate within this energy range is only $0.00019/\mathrm{yr}$ for UHECR interactions inside TXS0506+056.

\subsubsection*{Hadronic model}

\label{sec:hadronic}
\begin{figure}[t]
	\centering 
    \includegraphics[height=7cm]{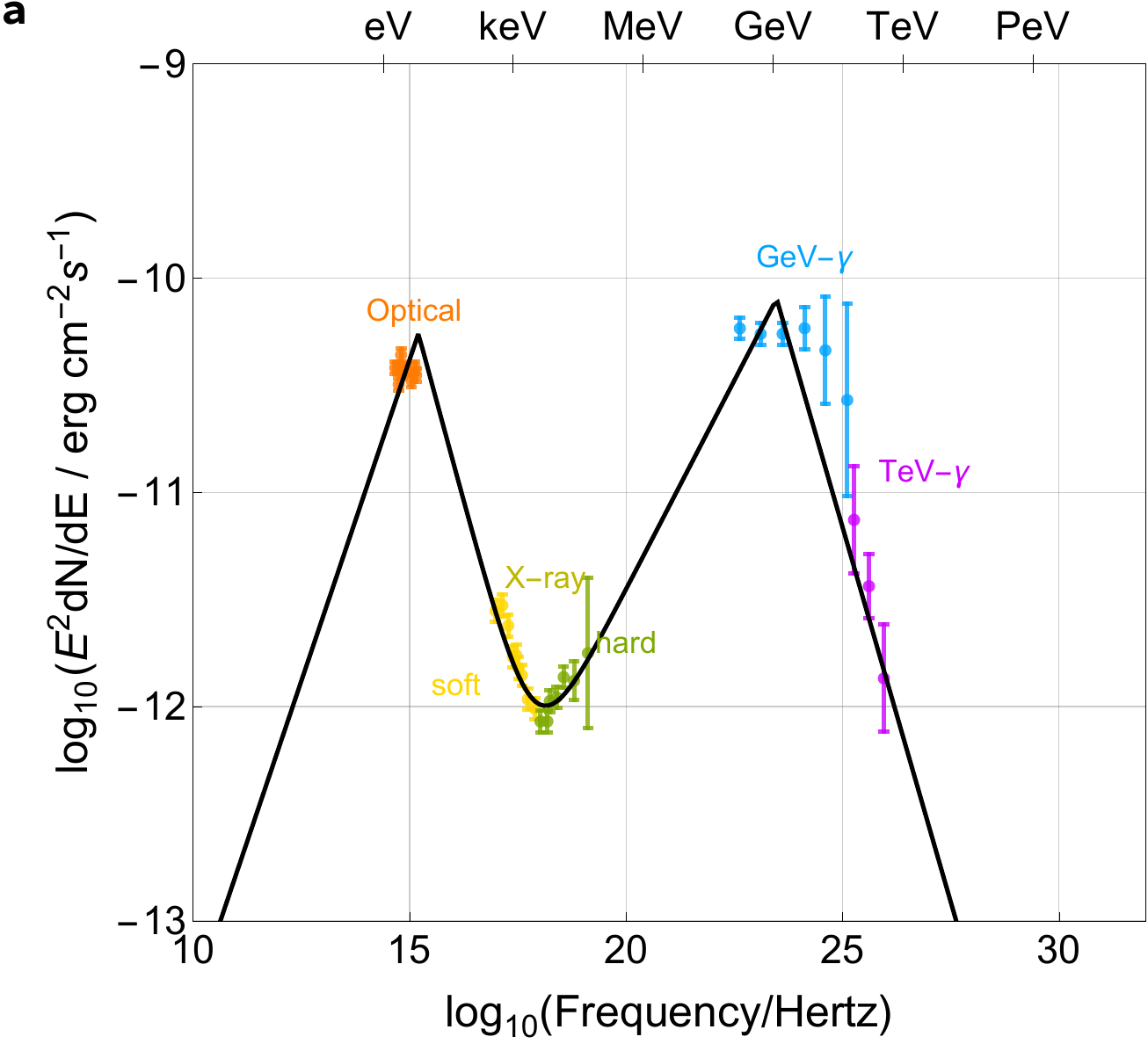}
    \includegraphics[height=7cm]{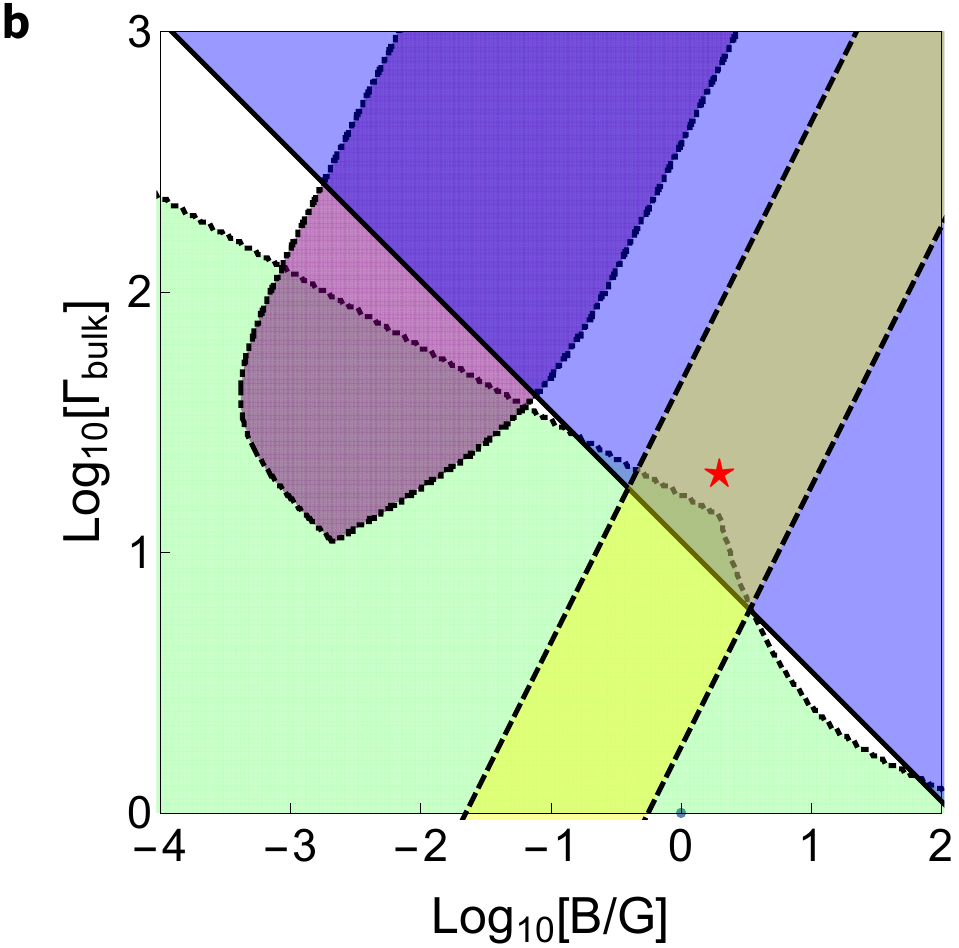}
	\caption{{\bf Extensive parameter-space scan with a semi-analytical approach for the hadronic model.} a) Four-power-law approximation of the spectral features used for an efficient scan of the hadronic parameters. b) Allowed regions for the model parameters in which the high-energy $\gamma$-ray radiation originates from the p$\gamma \to \pi^0,\pi^{\pm} \to {\rm e.m.  \, \, cascades}$ for one choice the blob-radius, $R_\mathrm{blob}^\prime=10^{16}~\mathrm{cm}$, and neutrino-energy, $E_{\nu}^\mathrm{ob}=250~\mathrm{TeV}$. The green area with dotted boundary corresponds to constraint (1); the blue upper-right region limited by the solid line refers to constraint (2); the yellow area between the dashed parallel lines reflects constraint (3); the violet lobe formed by dot-dashed boundaries represents constraint (4); see main text for details on these constraints.
}
	\label{fig:semianalytical}
\end{figure}

The {\it hadronic} model is defined as the scenario in which the low-energy part of the SED is produced by synchrotron emission of the primary electrons in the jet, whereas the high-energy component arises from hadrons via the process $p\gamma\rightarrow\pi^{0}\rightarrow\gamma\gamma$ and through synchrotron emission from secondary electrons generated through the reaction chain $p\gamma\rightarrow\pi^{\pm}\rightarrow\mu^{\pm}\rightarrow e^{\pm}$. If the target-photon density is high, as is the case here, the gamma rays induce electromagnetic cascades via pair-production and annihilation, $\gamma\gamma\rightarrow e^{\pm}\rightarrow \gamma \dots$. This is the most neutrino-optimistic scenario of blazar models in which comparable luminosities of gamma rays and neutrino are expected, since here the hadrons deposit a comparable share of their energy in neutrinos and photons.  

In the main text we describe why this class of neutrino-optimistic models is not applicable to TXS0506+056. Alternatives are sought by extensively scanning and constraining the parameter space in a semi-analytical analysis of the spectrum based on the method described in Appendix~A of \cite{2017ApJ...843..109G}. The following procedure yields contours for the allowed parameter regions: (1) approximate the entire SED by four power-law spectra (see \efigu{semianalytical}a); (2) choose a blob radius, $R_\mathrm{blob}^\prime$, and an observed neutrino energy, $E_{\nu}^\mathrm{ob}$; (3) vary the comoving magnetic field strength, $B^{\prime}$, and the Doppler factor of the blob, $\Gamma_\mathrm{bulk}$, on a 2D grid according to the constraints described in the next paragraph; (4) repeat this procedure for each combination of the parameters, $R_\mathrm{blob}^{\prime} \otimes E_{\nu}^\mathrm{ob}$ (in the present case $ 10^{15}~\mathrm{cm}<R_\mathrm{blob}^{\prime}<10^{19}~\mathrm{cm}$ and $10^{2}~\mathrm{TeV}<E_{\nu}^\mathrm{ob}<10^{3}~\mathrm{PeV}$). 

The {\it hadronic} model requires the following constraints to be met: (1) the synchrotron radiation of protons must not be brighter than the observed emission; (2) the inverse-Compton up-scattering of synchrotron photons may not dominate the high-energy emission; (3) the synchrotron emission from the hadronic secondaries should peak at $\nu_{\mathrm{peak},2}\sim10^{23}$ Hz, as observed (the width of the yellow band is related to the width of $\nu_{\mathrm{peak},2}$); (4) the emission of e$^\pm$ pairs from the electromagnetic process $p\gamma \to e^+ e^- p$ (Bethe-Heitler) must not exceed that observed in the X-ray band.

\efigu{semianalytical}b clearly demonstrates the absence of an overlap of all the four allowed regions for a specific choice of $R_\mathrm{blob}^{\prime}$ and $E_{\nu}^\mathrm{ob}$. The strongest constraints are imposed by the compatibility of X-ray data with the predicted emission following the Bethe-Heitler pair-production process (illustrated by the violet region in \efigu{semianalytical}b). Repeating this analysis for all combinations of $R_\mathrm{blob}^{\prime} \otimes E_{\nu}^\mathrm{ob}$, we always find a negative result and hence exclude the hadronic model as a possible explanation for the emission spectrum of TXS0506+056.  

\subsubsection*{Proton-synchrotron model}
\begin{figure}[t]
    \begin{center}
    \includegraphics[height=8cm]{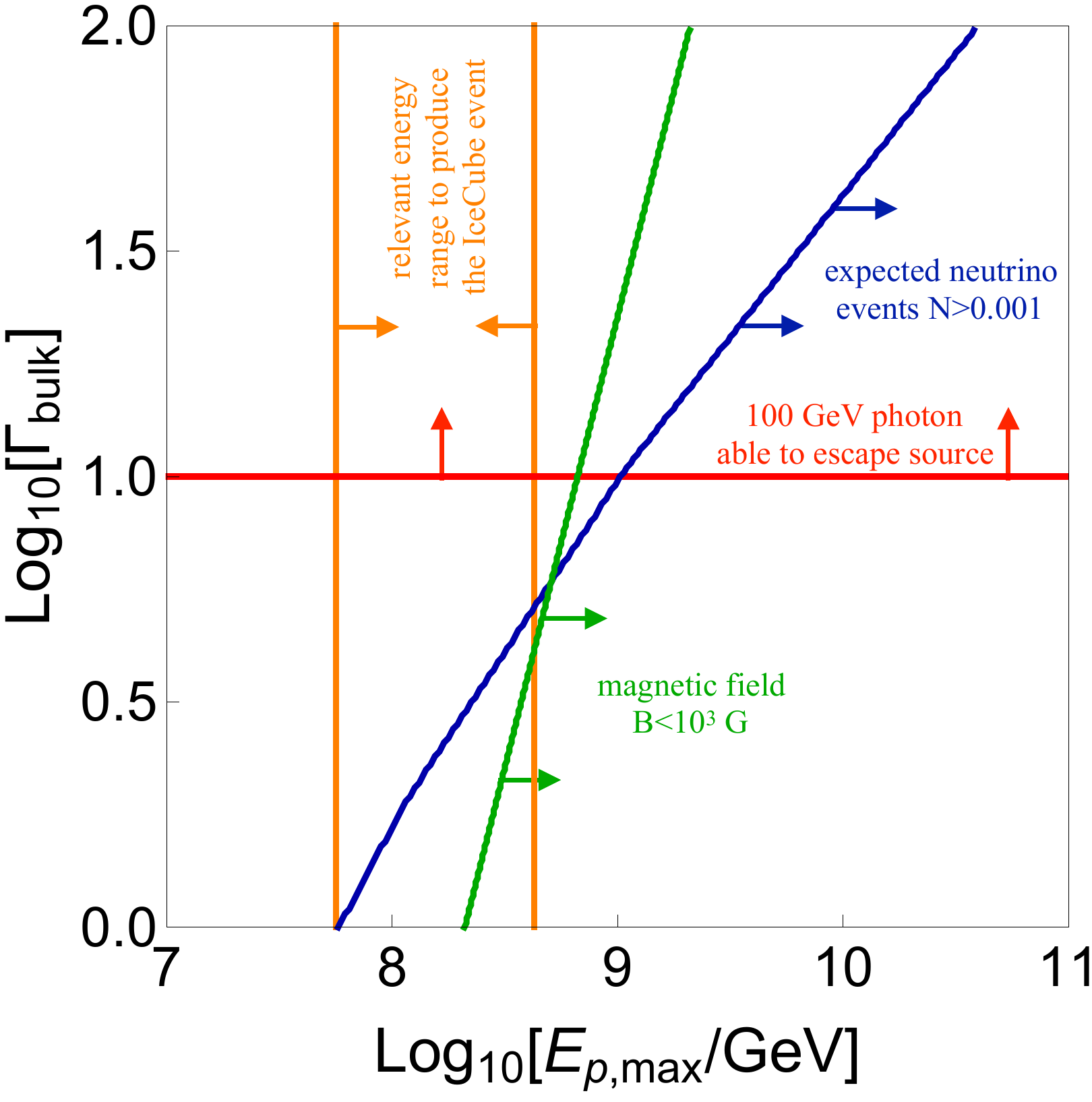}
    \end{center}
    \caption{\label{fig:psyn} {\bf Conditions to be met by a viable proton synchrotron model.} 
    The colored lines are allowed parameters ranges for the maximal proton energy, $E_{\rm p, max}$, and the Doppler factor, $\Gamma_{\rm bulk}$. The arrows point towards the allowed region. The {\it orange} region is restricted by the observed deposited energy, {\it red} by the presence of TeV gamma-rays, {\it green} by requiring a non-excessive magnetic field and {\it blue} by the probability to detect a neutrino in IceCube during an assumed 90 days flare period. A higher number of expected neutrinos moves the {\it blue} region to the right. The blob size is fixed to $R_\mathrm{blob}^\prime=10^{16}~\mathrm{cm}$.}
    \end{figure}
Another possibility to explain the second hump of the SED involves synchrotron emission of protons, hence the name proton-synchrotron model \cite{1992AA...253L..21M}. We use an analytical approach, similar to the analysis of the hadronic model.

Here we assume that monoenergetic protons are injected at a characteristic energy $E_{p}^{\rm ob}$ (observer's frame). By requiring that the proton-synchrotron spectrum reproduce the observed energy, $E_{\rm psyn}^{\rm ob}$, and flux of peak emission, $F_{\rm psyn}^{\rm ob}$, the energy densities of protons in the radiating blob, $u_{p}^{\prime}$, and magnetic field, $u_{B}^{\prime}$, can be expressed as functions of the two parameters blob radius, $R_{\rm blob}^{\prime}$, and Doppler factor, $\Gamma_{\rm bulk}$. For the $p\gamma$ interaction we adopt the $\Delta-$resonance approximation. The target photon energy is computed by the threshold condition of $p\gamma$ interactions $\sqrt{s} \sim m_\Delta \sim 1.2$ GeV. The target photon density is computed from a simplified SED as in \efigu{semianalytical}. The neutrino peak energy, $E_{\nu}^{\rm ob}$, and flux, $F_{\nu}^{\rm ob}$, are subsequently computed, taking into account synchrotron cooling of charged pions and muons.

To be acceptable, the model has to reproduce the observed spectral distribution and brightness of TXS0506+056, the presence of TeV gamma rays, and a peak neutrino energy in the range $183$--$4300$ TeV. \efigu{psyn} shows allowed ranges for the parameters $E_{p}^{\rm ob}$ and $\Gamma_{\rm bulk}$, given the constraints from the conditions above. The parameter scan demonstrates that no viable proton-synchrotron model produces at the same time neutrinos in the correct energy range and TeV gamma rays. We conclude that this class of models characteristically yields either detectable neutrino fluxes at excessively high energies (EeV range) or a very low neutrino flux at energies compatible with the current observation. Therefore, it is unlikely that a proton-synchrotron scenario can explain the neutrino coincidence with a gamma-ray flare of TXS0506+056.
\clearpage
\subsection*{Parameter table}

\begin{table*}[h]
    \centering
    \begin{tabular}{l|l|c|cc|c|c}
        \hline
        Param. & Description & Fit &  \multicolumn{2}{|c|}{Hybrid} & Hadronic & Leptonic \\
        \hline & & & Quiescent & Flare & Flare & Flare \\
        \hline
        $ z $ & Redshift &  fixed & \multicolumn{2}{c|}{$0.34$} & {$0.34$} & {$0.34$} \\
        $B^{\prime}$  & Magnetic field (G) & & {$0.007$} & {$0.14$} & $2.0$ & {$0.16$}\\
        $R_{\mathrm{blob}}^{\prime} $ & Blob radius (cm) && {$10^{17.5}$} & {$10^{16}$} & $10^{16}$ & {$10^{16}$} \\
        $\Gamma_\mathrm{bulk}$ & Doppler factor & & \multicolumn{2}{c|}{$28.0$} & $20.0$ & ${28.0}$ \\
        $L_{e,\mathrm{inj}}^{\prime}$  & $e^{-}$ injection luminosity (erg/s) & & $10^{40.5}$ & $10^{40.9}$ & $10^{41.3}$ & {$10^{41.0}$} \\
        $\alpha_{e}$ & $e^{-}$ spectral index & & {$-2.5$} & {$-3.5$} & $-2.3$ & $-3.5$ \\
        $\gamma_{e,\mathrm{min}}^{\prime}$ & Min. $e^{-}$ Lorentz factor & & \multicolumn{2}{c|}{$10^{4.2}$}  & $10^{3.3}$ &$10^{4.1}$ \\
        $\gamma_{e,\mathrm{max}}^{\prime}$ & Max. $e^{-}$ Lorentz factor & & {$10^{5.6}$} & {$10^{5.1}$} & $10^{4.4}$ &{$10^{5.9}$}\\
        $L_{p,\mathrm{inj}}^{\prime}$  & $p$ injection luminosity (erg/s)& & $10^{44.5}$ & $ 10^{45.7}$ & $ 10^{47.0} $ & -- \\
        $\gamma_{p,\mathrm{min}}^{\prime}$ & Min. $p$ Lorentz factor  & fixed & \multicolumn{2}{c|} {$10.0$} & $10.0$ & --\\
        $\gamma_{p,\mathrm{max}}^{\prime}$ & Max. $p$ Lorentz factor  & & \multicolumn{2}{c|} {$10^{5.4}$} & $10^{5.6}$ & --\\
        $\alpha_{p}$ & $p$ spectral index  &  fixed & \multicolumn{2}{c|} {$-2.0$} & $-2.0$ & --\\
        $\eta_\mathrm{esc}$ & escape velocity of $e^{\pm}$ and $p$& fixed & $c/300$ & $c/300$ & $c/10$ & $c/10$\\ 
        \hline
        Results & & & & & \\
        \hline
        $L_\mathrm{Edd}$ & Eddington luminosity * (erg/s)& & \multicolumn{2}{c|}{$10^{47.8}$}  & {$10^{47.8}$} & {$10^{47.8}$}\\
        $L_\mathrm{jet}$ & jet physical luminosity (in $L_\mathrm{Edd}$) & & {$0.4$} & {$6.2$} & {$62.8$} & $10^{-4}$\\
        $E_{\nu,\mathrm{peak}}$ & peak energy of $\nu$ spectrum (TeV)& & \multicolumn{2}{c|}{250} & 330 & --\\
        $N_{\nu}/yr$ & Expected IceCube $\nu$ events &  & $10^{-3.8}$ & 0.27 & 9.8 & 0\\
        \hline
    \end{tabular}
    \caption{{\bf Parameters of the models discussed in the main text.} Primed quantities refer to the rest frame of the radiation zone (blob). A sizable neutrino rate requires the jet power to exceed the Eddington luminosity during the flare. *We assume a black-hole mass of $5\times10^{9}~M_{\odot}$, similar to that of the nearby AGN M87.
    }
    \label{parametertable}
\end{table*}

\end{appendix}
\clearpage

\end{document}